\documentclass[%
 reprint,
superscriptaddress,
 amsmath,amssymb,
 aps,prl,
]{revtex4-2}

\usepackage{graphicx}
\usepackage[caption=false]{subfig}
\usepackage{dcolumn}
\usepackage{xcolor}
\usepackage{bm}
\usepackage{hyperref}
\usepackage{tkz-euclide}
\usetikzlibrary{arrows.meta}
\usepackage{physics}
\usetikzlibrary{shapes.misc}
\usetikzlibrary{decorations.pathmorphing}
\usepackage{comment}

\begin{document}

\preprint{APS/123-QED}

\title{A Diagrammatic Method for Non-Gaussian Photon Correlations in Weakly Coupled Atomic Ensembles}


\author{YangMing Wang}
 \email{ywan8652@uni.sydney.edu.au}
 \affiliation{
 ARC Centre of Excellence for Engineered Quantum Systems, \\School of Physics, The University of Sydney, Sydney, NSW 2006, Australia
}
\affiliation{Sydney Quantum Academy, Sydney, NSW, Australia}
\author{Sahand Mahmoodian}%
 \email{sahand.mahmoodian@sydney.edu.au}
\affiliation{
 ARC Centre of Excellence for Engineered Quantum Systems, \\School of Physics, The University of Sydney, Sydney, NSW 2006, Australia
}
\affiliation{Institute for Photonics and Optical Sciences (IPOS), School of Physics, The University of Sydney, NSW 2006, Australia}

\date{\today}

\begin{abstract}
We develop a scattering-theory formalism and use it to predict that a resonantly driven atomic ensemble weakly coupled to an optical mode can generate light with non-Gaussian correlations. Our approach---based on a perturbative diagrammatic expansion of multi-photon interactions---shows that photon-photon interaction mediated by the emitters causes the transmitted light to have a connected third-order correlation function $g_c^{(3)}$ containing non-Gaussian field correlations. The temporal pattern of $g_c^{(3)}$ is explained using the microscopic transport processes in terms of our diagrams. We derive an optical-depth-dependent bound on the perturbative truncation error and benchmark our predictions against cascaded master-equation and quantum trajectory simulations for small ensembles with $\mathrm{OD}\leq 2$, identifying a regime of controlled accuracy under weak driving. Our predictions can be experimentally demonstrated in state-of-the-art nanofibre-coupled atomic ensembles.
\end{abstract}
\maketitle
{\it Introduction.}---Photon-photon interactions in optically driven atomic ensembles can alter the quantum state of an incident laser field. The interplay of nonlinear interactions, due to the atoms in the ensemble, and the elastic scattering of laser light out of the ensemble allows manipulating the quantum nature of light \cite{Mahmoodian2018, RevModPhys.95.015002}. For example, photons can become strongly correlated, modifying both intensity correlations \cite{Prasad2020, Cordier2023, Ferioli2025PRLa} and electric field fluctuations \cite{Mahmoodian2021, Lu1998PRL}. Such works have largely focused on quantifying two-body photon correlations. Three-photon correlations have been measured with some atomic ensemble systems \cite{Stiesdal2018, Liang2018Science, Ornelas-Huerta2021PRL, Das2025arXiv}, but whether atomic ensembles can generate non-Gaussian light remains an open question. By non-Gaussian field correlations, we mean that the Wigner function of the output state is non-Gaussian in phase space, such that its higher-order field moments are not, in general, determined by the first and second moments alone.

The non-Gaussianity of a state of light can be probed by measuring the cumulants of its intensity correlation functions. However, two-body correlations alone can give rise to a nonzero connected third-order intensity correlation function, $g_c^{(3)}$~\cite{Cardin2024,Hackl2021}. This raises a question: can three-photon correlations be generated in weakly coupled atomic ensembles, survive photon loss due to scattering, and provide information about the field’s non-Gaussianity through $g_c^{(3)}$? Until now, a tractable theoretical framework for answering this questions has remained elusive. Previous work has used the Maxwell-Bloch equations under mean-field approximations \cite{McCall1969PR, Hammerer2010RMP, Goncalves2025PRXQ, Kusmierek2025}, which typically cannot capture quantum correlations. Beyond the mean-field level, one can employ a complete cascaded master equation \cite{Gardiner1993PRL, Carmichael1993PRL}, however, for large ensembles the direct computation of higher-order correlation functions is challenging, even if employing cumulant-expansion \cite{Mahmoodian2023, Plankensteiner2022Quantum} or matrix-product operator \cite{manzoni2017NCOMM, Mahmoodian2020PRX} methods. 

\begin{figure}
  \centering
  \begin{tikzpicture}
 
    \node[inner sep=0] (main) {\includegraphics[width=0.5\textwidth,height=1.4cm]{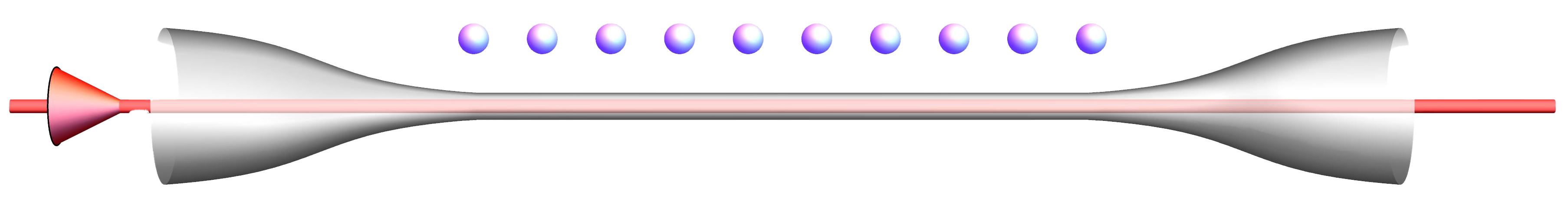}};
   
    \node[anchor=north east, xshift=9.36cm, yshift=1.2cm] at (main.north west) {%
      \begin{tikzpicture}[scale=0.5]
        \draw[->, thick, >={Stealth[length=3mm, width=2mm]}] (0.2,-1.3) -- (0.2,0) node[above] {$(1-\beta)\Gamma_{\rm tot}$}
    (0.2,-1.3) -- (0.2,0) node[above] {$(1-\beta)\Gamma_{\rm tot}$};
        \draw[->, thick, >={Stealth[length=3mm, width=2mm]}]
    (0.2,-2) to[out=-90, in=180] node[below] {$\beta\Gamma_{\rm tot}$} (2.2,-2.7) ;
        \node[anchor=west] at (-7.5,-4) {$\ket{\alpha}$};
        \node[anchor=west] at (8.3,-4) {$\ket{\rm out}$};
        
        \draw[dashed,thick] (-2.8,-1.5)--(-4.4,0.4);
        \draw[dashed,thick] (-3.1,-1.8)--(-4.4,-0.6);
        \draw[thick] (-4.6,0.4)--(-5.3,0.4) node[left,xshift=-4pt] {$\ket{e}$};
        \draw[thick] (-4.6,-0.6)--(-5.3,-0.6) node[left,xshift=-4pt] {$\ket{g}$};
        
        \draw[thick] (-4.95,-0.1) circle [radius=0.9];
       
        \fill[black] ($(-4.6,-0.6)!0.5!(-5.3,-0.6)$) circle (3pt);
      \end{tikzpicture}
    };
  \end{tikzpicture}
  \caption{\label{fig:3d plot} An array of $M$ chirally coupled two-level atoms (blue circles) resonantly driven by an external coherent field $\ket{\alpha}$ producing a strongly correlated output photon state $|\text{out}\rangle$. Each atom couples to the waveguide with rate $\beta\Gamma_{\text{tot}}$ and to its own independent external loss channel with rate $(1-\beta)\Gamma_{\text{tot}}$. }
\end{figure}
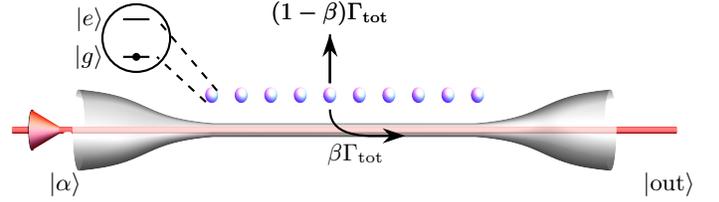

In this work, we propose an analytic and diagrammatic framework to calculate up to three-photon transport through an atomic ensemble. Using the Maxwell-Bloch approach,  the ensemble can be considered to be an array of $M$ atoms unidirectionally coupled to an optical nanofibre as shown in Fig.~\ref{fig:3d plot}. The coupling efficiency between each atom and waveguide is weak $\beta \ll 1$, but $M\gg1$. Our diagrammatic approach takes advantage of the weak coupling and uses $\beta$ as an expansion parameter, enabling us to analytically compute leading-order contributions to the three-photon-correlated component of the wavefunction and the connected third-order correlation function $g_c^{(3)}$. To show the non-Gaussian nature of the field, we decompose $g_c^{(3)}$ into two contributions, \(g_c^{(3)}=\widetilde{g}_c^{(3)}+\Delta g_c^{(3)}\). The first, $\widetilde{g}_c^{(3)}$, contains contributions expressible in terms of products of first- and second-order moments of the field operators, according to Isserlis' theorem for Gaussian states. The remainder, $\Delta g_c^{(3)}$, vanishes for any Gaussian state; hence, $\Delta g_c^{(3)}\neq0$ provides a sufficient condition for non-Gaussianity. Our diagrammatic analysis further shows that $\Delta g_c^{(3)}$ is dominated by wavefunction components arising from connected three-photon interactions.

{\it Model.}---The model for our system is depicted in Fig.~\ref{fig:3d plot}, where $M$ atoms are unidirectionally coupled to a propagating channel in a waveguide. Here we have illustrated the example of a nanofibre geometry where atoms are trapped near the fibre and interact via its evanescent field. Each atom couples to the propagating channel at a rate $\Gamma = \beta\Gamma_{\rm tot}$ and to its individual loss channel at a rate $\gamma = (1-\beta)\Gamma_{\rm tot}$, where $\beta$ quantifies the coupling efficiency and $\Gamma_{\rm tot}$ is the overall atomic decay rate. The emitters are driven by a resonant coherent field from one end of the waveguide. The emitter spacing exceeds one resonant wavelength of the two-level atoms which ensures negligible collective emission outside the waveguide. Typical experimental values for atomic ensembles coupled to nanofibres yield $\beta \sim 0.01$~\cite{Prasad2020}. Despite the small individual coupling strength $\beta$, the OD of the entire atomic array becomes significant $\mathrm{OD}=4\beta M=\mathcal{O}(1) $, where $\mathcal{O}(\cdot)$ indicates big-O notation. This enables the collective nonlinear effects and strong dissipation to substantially modulate photon transport properties, which, as we show, results in strongly correlated photons and non-Gaussian output states of light.

For compactness, we define a Hamiltonian functional (with $\hbar = c = 1$),
\begin{eqnarray}
    \mathcal{H}_m[\hat{c},\chi] &=& \int_{-\infty}^{\infty} d x  
    \hat{c}^\dagger(x) (-i\partial_x) \hat{c}(x) 
    + \delta(x)\sqrt{\chi}\big[ \hat{\sigma}_m^+ \,\hat{c}(x)\nonumber\\
    &+& {\rm h.c.}\big].
\end{eqnarray}
The local scattering event at the $m$th atom is described by a one-dimensional two-channel model, where each atom interacts with the waveguide and a loss channel. The Hamiltonian of this model reads $\hat{H}_m=\mathcal{H}_m[\hat{a},\Gamma]+\mathcal{H}_m[\hat{b}_m,\gamma]$, where $\hat{a}^\dagger(x)$ and $\hat{b}_m^\dagger(x)$ create photons in the propagating and $m$th atom's loss channels respectively, while $\hat{\sigma}_m^\pm$ are Pauli operators for the $m$th two-level emitter. Then, by applying the following Weyl transformation,
$\hat{a}_m^{(e)}(x) =\sqrt{\beta}\,\hat{a}(x) + \sqrt{1-\beta}\,\hat{b}_m(x),
\hat{a}_m^{(o)}(x) = \sqrt{1-\beta}\,\hat{a}(x) - \sqrt{\beta}\,\hat{b}_m(x)$,
$\hat{H}_m$ splits into two decoupled even and odd channels $\hat{H}_m = \hat{H}_m^{(e)} + \hat{H}_m^{(o)}$. The odd channel Hamiltonian $\hat{H}_m^{(o)} = \int_{-\infty}^{\infty} dx\, \hat{a}_m^{\dagger(o)}(x) \bigl(-i\partial_x\bigr) \hat{a}_m^{(o)}(x)$ describes non-interacting transport. All the information about atom-photon and photon-photon interactions at the $m$th atom is encoded in the even channel Hamiltonian $\hat{H}_m^{(e)}=\mathcal{H}_{m}[\hat{a}_{m}^{(e)},\Gamma_{\rm tot}]$. We note that $\beta$ only enters the dynamics through the Weyl transformation and not the Hamiltonian itself.

The $n$-photon scattering in the even subspace can be fully described by the $S$-matrix derived from Bethe's Ansatz~\cite{Yuds1985, Rupasov1984JETP}. The full $S$-matrix element encodes all the information of free propagation of photons, photon-atom and photon-photon interactions. We refer to free propagation of photons and individual elastic photon-atom scattering as \textit{disconnected processes} because they factorize into independent terms in the $S$-matrix. The \textit{connected part} of the $S$-matrix is defined by subtracting all the products of the disconnected processes from the full $S$-matrix, enabling us to extract the genuine effects of $n$-body photon-photon interactions~\cite{WeinbergQFT}. The expression of the full and connected two- and three-photon $S$-matrix is given in the Supplemental Material~\cite{SM}.

{\it Computational Procedure}---Because photon transport is chiral, scattering at a downstream atom does not affect scattering at upstream atoms. The array can therefore be treated sequentially, with the propagating output from the $m$th atom providing the input to the $(m+1)$th atom.  The local scattering event at the $m$th atom can be calculated by the following steps: (1) Transform the incoming state from the propagating channel into even/odd channel picture, where the input from the loss channel is vacuum. (2) Multiply the term with $n$ even photons by the $n$-photon $S$-matrix. (3) Transform back to propagating/loss channel picture. For single-photon scattering in the even channel, energy conservation and unitarity imply that the $S$-matrix is a frequency-dependent phase factor. Solving the regularized time-independent Schrödinger equation gives the phase $s_k^{(e)}=(k-i\Gamma_{\rm tot}/2)/(k+i\Gamma_{\rm tot}/2)$~\cite{Yudson2008}. Combining this with the unchanged odd channel and transforming back to the propagating/loss basis yields the transmission and loss coefficient $t_k=1-i\beta\Gamma_{\mathrm{tot}}/(k+i\Gamma_{\mathrm{tot}}/2)$,
$r_k=-\sqrt{\beta(1-\beta)}i\Gamma_{\rm tot}/(k+i\Gamma_{\rm tot}/2)$, 
respectively, with $k=0$ corresponding to resonance~\cite{SM}. 
Correlated two- and three-photon scattering is captured by the multiplications of their respective connected scattering matrices $\beta^2 \hat{S}^C_{p_1p_2,k_1k_2}$ and $\beta^3 \hat{S}^C_{p_1p_2p_3,k_1k_2k_3}$ if the participating photons remains in the waveguide after scattering. The powers of $\beta$ follow from Weyl transformation: each incoming photon interacting with the atom contributes $\sqrt{\beta}$, while each outgoing photon contributes $\sqrt{\beta}$ if transmitted or $\sqrt{1-\beta}$ if emitted into the loss channel.  The full output state is thus obtained by applying the local $S$-matrices successively to the input state, in order from the first atom to the last along the direction of photon propagation.

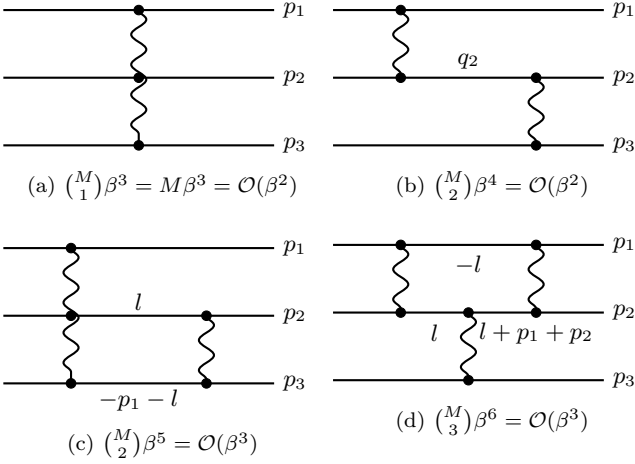
\begin{figure}[t]
  \centering
  \subfloat[$\binom{M}{1}\beta^3=M\beta^3= \mathcal{O}(\beta^2)$\label{fig:fork}]{%
    \begin{tikzpicture}[baseline=(current bounding box.center),scale=0.9]
      \draw[thick] (-2, 1) -- (2, 1) node[right]{$p_1$}; 
      \draw[thick] (-2, 0) -- (2, 0) node[right]{$p_2$}; 
      \draw[thick] (-2, -1) -- (2, -1) node[right]{$p_3$}; 
      
      \draw[thick, decorate,
        decoration={
          snake,
          amplitude=1mm,
          segment length=4mm
        }] (0, 1) -- (0, -1);
      
      \filldraw[black] (0, 0) circle (2pt); 
      \filldraw[black] (0, 1) circle (2pt); 
      \filldraw[black] (0, -1) circle (2pt); 
    \end{tikzpicture}
  }
  \hfill
  \subfloat[$\binom{M}{2}\beta^4= \mathcal{O}(\beta^2)$\label{fig:crank}]{%
    \begin{tikzpicture}[baseline=(current bounding box.center),scale=0.9]
      \draw[thick] (-2, 1) -- (2, 1) node[right]{$p_1$}; 
      \draw[thick] (-2, 0) -- (2, 0) node[right]{$p_2$}; 
      \draw[thick] (-2, -1) -- (2, -1) node[right]{$p_3$}; 
      
      \draw[thick, decorate,
        decoration={
          snake,
          amplitude=1mm,
          segment length=4mm
        }] (-1, 1) -- (-1, 0);
      \draw[thick, decorate,
        decoration={
          snake,
          amplitude=1mm,
          segment length=4mm
        }] (1, 0) -- (1, -1);
      
      \filldraw[black] (-1, 0) circle (2pt); 
      \filldraw[black] (1, 0) circle (2pt);
      \filldraw[black] (-1, 1) circle (2pt); 
      \filldraw[black] (1, -1) circle (2pt); 
      \node at (0,0.25) {$q_2$};
    \end{tikzpicture}
  }
  
  \subfloat[$\binom{M}{2}\beta^5=\mathcal{O}(\beta^3)$\label{fig: three two diag}]{%
    \begin{tikzpicture}[baseline={([yshift=5pt]current bounding box.center)},scale=0.9]
      \draw[thick] (-2, 1) -- (2, 1) node[right]{$p_1$}; 
      \draw[thick] (-2, 0) -- (2, 0) node[right]{$p_2$}; 
      \draw[thick] (-2, -1) -- (2, -1)node[right]{$p_3$}; 
      \draw[thick, decorate,
        decoration={snake, amplitude=1mm, segment length=4mm}]
        (-1, 1) -- (-1, -1);
      \draw[thick, decorate,
        decoration={snake, amplitude=1mm, segment length=4mm}]
        (1, 0)-- (1, -1);
      
      \filldraw[black] (-1, -1) circle (2pt);
      \filldraw[black] (-1, 0) circle (2pt); 
      \filldraw[black] (1, 0) circle (2pt);
      \filldraw[black] (-1, 1) circle (2pt); 
      \filldraw[black] (1, -1) circle (2pt);  
      \node at (0, 0.2) {$l$};
      \node at (0,-1.25) {$-p_1-l$};
    \end{tikzpicture}
  }
  \hfill
    \subfloat[$\binom{M}{3}\beta^6= \mathcal{O}(\beta^3)$\label{fig:twotwotwo mid diag}]{%
    \begin{tikzpicture}[baseline=(current bounding box.center) ,scale=0.9]
      \draw[thick] (-2, 1) -- (2, 1) node[right]{$p_1$};
      \draw[thick] (-2, 0) -- (2, 0) node[right]{$p_2$};
      \draw[thick] (-2, -1) -- (2, -1) node[right]{$p_3$};
      
      \draw[thick, decorate,
        decoration={snake, amplitude=1mm, segment length=4mm}]
        (-1, 1) -- (-1, 0);
      \draw[thick, decorate,
        decoration={snake, amplitude=1mm, segment length=4mm}]
        (1,1) -- (1,0);
      \draw[thick, decorate,
        decoration={snake, amplitude=1mm, segment length=4mm}]
        (0, 0) -- (0, -1);
      
      \filldraw[black] (-1, 0) circle (2pt); 
      \filldraw[black] (-1, 1) circle (2pt); 
      \filldraw[black] (0, -1) circle (2pt);
      \filldraw[black] (0, 0) circle (2pt);
      \filldraw[black] (1, 0) circle (2pt);
      \filldraw[black] (1, 1) circle (2pt); 
      \node at (0,0.7) {$-l$};
      \node at (-0.5,-0.3) {$l$};
      \node at (1.0,-0.3) {$l+p_1+p_2$};
    \end{tikzpicture}
  }
  \caption{\label{fig:diagrams} Diagrammatic representation of the leading photon transport processes contributing to the non-Gaussianity of outgoing light. (a) The diagram including one three-photon interaction is called a 3-vertex diagram, and (b) the diagram including two two-photon interactions is called 4-vertex diagram. The momenta of the photons after the interaction are labeled above the horizontal lines. Lines without momentum labels correspond to resonant photons. The order estimates below each diagram show their respective scaling at large optical depths.}
\end{figure}

{\it Diagrams}---Expressing the interacting scattering in terms of the connected part of the $S$-matrix clarifies the way in which connected photon–photon interactions scale with the various powers of coupling strength~$\beta$. Since $\beta \ll1$, this scaling behavior motivates us to perturbatively treat the effect of photon-photon interactions in terms of the order in $\beta$. If a three-photon resonant transport process has no photon-photon interaction, the outgoing state is simply the incoming state of the array multiplied by $t_0^{3M}$. Transport processes generating three-photon correlations are the main interest of this work. In Fig.~\ref{fig:diagrams}, we sketch the interaction processes where three-photon trajectories are connected by interaction lines. These diagrams are drawn according to the following rules: (1) Each horizontal line indicates the trajectory of each photon being transported through the array from the left to the right. The individual elastic atom-photon scatterings are not explicitly indicated on the line. (2) Each black circular dot on the photon trajectory represents an atomic site where the photon interacts with other photons. The dots on the same column corresponds to the same atomic site. (3) A $n$-photon interaction is represented by a wavy line connecting $n$ dots. (4) The outgoing momenta after transports are denoted as $p_1, p_2, p_3$ at the right end of the diagram. We note that black dots can be any atom in the array, therefore each diagram in Fig.~\ref{fig:diagrams} represents the family of all possible transport processes with the interactions happening on different atomic sites. For example, the 3-vertex diagram in Fig.~\ref{fig:diagrams}(a) shows three photons interacting via a single atom during the transport through $M$ atoms. This can occur at any atom within the ensemble and therefore the amplitude of this type of processes has a combinatorial factor of $M$. The 4-vertex diagram in Fig.~\ref{fig:crank} displays a transport process where a pair of photons interact at one atom while the third photon scatters off the atom independently. Subsequently, a different pair of photons interact at another downstream atom.

{\it Order Estimation}---We identify leading-order diagrams by estimating each diagram's amplitude. Each black-dot vertex associated with an interaction between an atom and a transmitted photon contributes a factor of $\beta$ to the amplitude. In addition to $\beta$ contributing to a diagram's order, we must also consider the combinatorics of the process due to interactions potentially occurring at any atom in the ensemble. The combinatorics of a diagram with photon-photon interactions on $n$ atoms is $\binom{M}{n}$. For larger OD $\beta M=\mathcal{O}(1)$, the order of magnitude of a diagram is estimated by multiplying $\beta$ and the combinatorial factor. For example, the order of the 4-vertex diagram, Fig.~\ref{fig:crank}, is $\binom{M}{2}\beta^4\approx M^2 \beta^4= \mathcal{O}(\beta^2)$. For the order estimation in small or intermediate OD,  because the power of $\beta$ is fixed for each diagram, the amplitude of the diagram grows monotonically with $M$, therefore the leading diagrams at small or intermediate OD are a subset of those at large OD. Hence, perturbative results derived by identifying leading-order diagrams in large-OD regime remain valid, and gain accuracy, when applied to lower ODs.

At large OD, 3- and 4-vertex diagrams are the leading-order three-photon connected diagrams whose amplitudes are denoted as $\hat{T}^{3v}\psi_0$ and $\hat{T}^{4v}\psi_0$. Their expression derived based on diagrammatic rules is given in~\cite{SM}. They are both at $\mathcal{O}(\beta^2)$, which is the leading-order contribution among all possible three-photon-connected diagrams. The subleading order $\mathcal{O}(\beta^3)$ three-photon connected diagrams includes two types of diagrams: (1) the diagrams with one three-photon interaction and one two-photon interaction, and (2) the diagrams with three two-photon interactions. One representative of each type is shown in Figs.~\ref{fig: three two diag} and ~\ref{fig:twotwotwo mid diag}. The subleading order diagrams are negligible at $\beta\approx 1\%$, but make modest quantitative corrections on top of the leading-order diagrams at large OD with $\beta\approx 5\%$. As the diagrams feature a loop structure, we refer to these as loop-order diagrams  and their explicit calculation is presented in our accompanying work~\cite{WDM2025}. To assess the accuracy of truncating three-photon transport at loop order, we derive a loose upper bound on the $L^2$-norm of the wavefunctions from omitting higher-order diagrams~\cite{SM}. We compare this bound against $\|\hat{T}^{3v}\psi_0\|_2+\|\hat{T}^{4v}\psi_0\|_2$, using the sum of the individual leading-order norms to avoid phase cancellations between the two amplitudes. This ratio grows with OD, but it suggests that the leading-order diagrams significantly dominates over the total contribution of higher-order diagrams when $\beta\leq 5\%$ and OD $\leq 3.5$. For $\beta\lesssim 2\%$, higher-order diagrams are quantitatively negligible.

\begin{figure}[t]
    \centering
    \includegraphics[width=\linewidth]{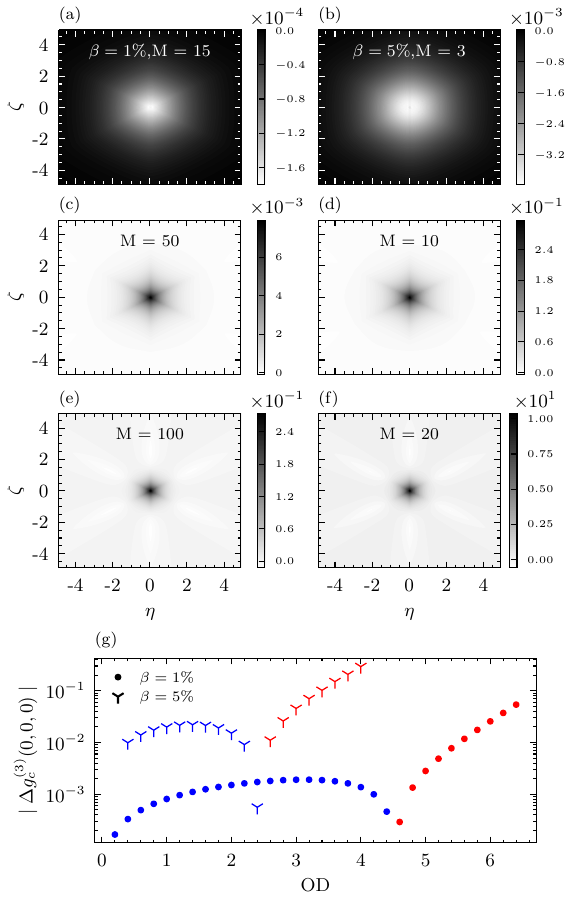}
    \caption{\label{fig:g3c} Connected third-order correlation function with various OD $g_c^{(3)}(R,\eta,\zeta)$ in Jacobi coordinates, with center of mass $R=0$. The six-fold symmetry reflects three symmetry axes corresponding to two-photon coincidences. $g_c^{(3)}$ is computed at tree-level diagram for $\beta=1\%$ while the loop order correction is added for $\beta= 5\%$. In (g), the deviation of from Isserlis's factorization $\Delta g_c^{(3)}(0,0,0)$ is plotted against OD. Blue (red) markers indicate a negative (positive) sign. }
\end{figure}

{\it Connected third-order correlation function and non-Gaussianity}---To connect the three-photon scattering theory with experiment, we now consider an on-resonance coherent-state input. Under the weak-drive condition $\mathcal{O}(P_{\rm in}/\Gamma_{\rm tot})= \mathcal{O}(\beta)$ \footnote{Strictly, we require $\mathcal{O}(P_{\rm in}/\Gamma_{\rm tot})= \mathcal{O}(\beta^2)$, but it is shown in the accompanying work~\cite{WDM2025} that this condition can be loosened to be $\mathcal{O}(P_{\rm in}/\Gamma_{\rm tot})= \mathcal{O}(\beta)$.}, the connected third-order correlator can be expanded in $P_{\rm in}/\Gamma_{\rm tot}$, with the leading term determined by the three-photon transport processes above. The connected third-order correlation function is defined as
\begin{equation*}
  g_c^{(3)}(x_1, x_2, x_3) = 2 + g^{(3)}(x_1, x_2, x_3) - \sum_{i < j} g^{(2)}(x_i, x_j),
\end{equation*}
where $g^{(n)}(x_1,\dots,x_n)=\expval{:\Pi_{j=1}^n \hat{a}^\dagger(x_j)\hat{a}(x_j):}/\Pi_{k=1}^n \expval{\hat{a}^\dagger(x_k)\hat{a}(x_k)}$ is the $n$th-order intensity correlation function and $:\,\,:$ indicates normal ordering. For non-vanishing power, $g_c^{(3)}(x_1, x_2, x_3)$ vanishes when either one photon is far away from the other two~\cite{Lemonde2014PRA, Jachymski2016,Stiesdal2018}. Moreover, if the state of light has only Gaussian field correlation, both $g^{(3)}$ and $g^{(2)}$ can be determined by the second-order moment and mean of $\hat{a}(x)$ and $\hat{a}^\dagger(x)$ by Isserlis's theorem~\cite{Cardin2024,Shi2018,Hackl2021} (see~\cite{SM}). A non-vanishing deviation from Isserlis's factorization $\Delta g_c^{(3)}$ implies the state is non-Gaussian. Our theory produces the scattered two- and three-photon wavefunction denoted $\psi_2(x_1, x_2)=t_0^{2M}+\phi_2(x_1,x_2)$ and $\psi_3(x_1,x_2,x_3)=t_0^{3M}+t_0^M\sum_{i<j}\phi_2(x_i,x_j)+\phi_3(x_1,x_2,x_3)$ respectively. Here, $\phi_2$ and $\phi_3$ are the two- and three-photon correlated parts of the wavefunction. For weak driving, and at leading order, we have,
\begin{eqnarray} \label{eq:g3ng2}
g^{(3)}(x_1, x_2, x_3)
&\approx&|\psi_3(x_1, x_2, x_3)|^2/t_0^{6M}, \nonumber\\
g^{(2)}(x_1,x_2)&\approx& |\psi_2(x_1,x_2)|^2/t_0^{4M},
\end{eqnarray}
The output power $\expval{\hat{a}^\dagger(x)\hat{a}(x)}\approx P_{\rm in}t_0^{2M}$. We evaluate $\phi_2$ and $\phi_3$ analytically, but for brevity their expressions are presented in~\cite{WDM2025}. By using \eqref{eq:g3ng2} and Isserlis's theorem, we obtain \emph{the central link between three-photon connected scattering and non-Gaussianity}
\begin{eqnarray} \label{eq:Delta g3c analytic}
    & &\Delta g_c^{(3)}(x_1, x_2, x_3) \approx  \frac{2\phi_3(x_1,x_2,x_3)}{t_0^{5M}}\Big[t_0^{2M}  +\sum_{i<j}\phi_2(x_i,x_j) \Big]. \nonumber
\end{eqnarray}
It expresses the experimentally measurable deviation from Isserlis factorization directly in terms of the three-photon-connected scattering amplitude $\phi_3$.

In Fig.~\ref{fig:g3c} we plot the connected third‐order correlation $g_{c}^{(3)}(x_{1},x_{2},x_{3})$ using (\ref{eq:g3ng2}) for an atomic ensemble with $\beta=1\%$ and $5\%$. The value $\beta=1\%$ represents the approximate value from past experiments \cite{Prasad2020, Mahmoodian2021, Mahmoodian2023}, while $\beta=5\%$ is what could be obtained if the atoms are trapped more closely to the fibre and/or if the light-matter coupling is enhanced, for example, with slow light using fibre gratings. Because the outgoing steady state of light is translationally invariant, $g_c^{(3)}(x_1,x_2,x_3)$ depends only on two independent relative separations rather than on the absolute positions of the three photons. We therefore introduce Jacobi coordinates, which separate the center-of-mass motion from the relative three-photon configuration: $R=\frac{x_{1}+x_{2}+x_{3}}{\sqrt{3}},\quad
\eta=\frac{x_{1}-x_{2}}{\sqrt{2}},\quad
\zeta=\sqrt{\tfrac{2}{3}}\left(\frac{x_{1}+x_{2}}{2}-x_{3}\right)$
with $R=0$ fixed by translational invariance. Here, $\eta$ measures the separation within one photon pair, while \(\zeta\) measures the separation of the third photon from the center of that pair. In the $\eta$--$\zeta$ plane, the value of $g_c^{(3)}$ at the origin $(\eta,\zeta)=(0,0)$ corresponds to a three-photon coincidence $x_1=x_2=x_3$, i.e., enhanced (positive) or suppressed (negative) simultaneous detection of all three photons. Three axes $\eta=0$, $\zeta=\sqrt{3}/3 \eta$ and $\zeta=-\sqrt{3}/3 \eta$  corresponding to two-photon coincidence events with the third photon separated (except at their intersection (0,0)). $g_c^{(3)}$ is symmetric with respect to the three axes because of bosonic permutation symmetry. Furthermore, this three-fold symmetry is doubled to be a six-fold symmetry because $g_{c}^{(3)}$ does not distinguish whether a photon pair arrives before or after the lone photon. 

As the optical depth increases, $g_{c}^{(3)}$ evolves from being everywhere negative, through a regime of uniform positivity, to a characteristic pattern with a bright central peak surrounded by six negative ``outer legs.'' Under approximation~(\ref{eq:g3ng2}), 
\begin{eqnarray} \label{eq:g3c approx}
    & & g_c^{(3)}(x_1,x_2,x_3)\approx \Delta  g_c^{(3)}(x_1,x_2,x_3)\\
      &+& 2\sum_{(i,j,k)\in  A}\phi_2(x_i,x_j)\phi_2(x_j,x_k)/t_0^{4M}, \nonumber
\end{eqnarray}
 where $A= \{(1,2,3),(2,1,3),(1,3,2)\}$. The second term, is the Gaussian contribution $\widetilde{g}_c^{(3)}(x_1,x_2,x_3)$ given by Isserlis theorem. Thus, $g_c^{(3)}=\Delta g_c^{(3)}+\widetilde g_c^{(3)}$ separates into two physically distinct contributions: the non-Gaussian deviation $\Delta g_c^{(3)}$, which is controlled by the three-photon-connected scattering amplitude $\phi_3$, and the Gaussian contribution $\widetilde g_c^{(3)}$, which is constructed entirely from products of the correlated two-photon amplitudes $\phi_2$.

{\it Microscopic explanation for $g_c^{(3)}$ and $\Delta g_c^{(3)}$}---We explain the spatial pattern in Fig.~\ref{fig:g3c} in terms of the relative amplitude of various transport processes in Fig.~\ref{fig:diagrams} and two-photon interaction. Using the $\beta$-factor scaling established above from the Weyl transformation, each local connected $n$-photon interacting scattering, after which all photon remains in the waveguide, contributes a factor $\beta^n$ to the corresponding wavefunction amplitude. At low optical depth (top row of Fig.~\ref{fig:g3c}), $4\beta M=0.6$, $\phi_3$ is dominated by the process in 3-vertex Fig.~\ref{fig:fork}, which contains a single three-photon interaction at $\mathcal{O}(\beta^3)$ and is negative. By comparison, $\phi_2$ is dominated by a single two-photon interaction at $\mathcal{O}(\beta^2)$. The resulting $\phi_2\phi_2$ contribution to $g_c^{(3)}$ is at $\mathcal{O}(\beta^4)$ therefore small in this OD. As the optical depth increases (second row of Fig.~\ref{fig:g3c}), e.g., to $4\beta M=2$, more atom sites are available to accommodate photon-photon interactions. Sequential photon-photon interactions result in the accumulation of the amplitude of processes with positive sign in both the $\phi_3$ and $\phi_2\phi_2$. In $\phi_3$, 4-vertex diagram grows in importance because it is enhanced by the combinatorial factor $\binom{M}{2}$ counting the choice of two emitter sites. At the same time, the magnitude of $\phi_2$ itself becomes larger since the leading single two-photon interaction contribution is enhanced by a combinatorial factor $M$ from the $M$ possible sites at which the two-photon interaction can occur. At large OD (third row of Fig.~\ref{fig:g3c}) $4\beta M=4$, $\phi_2\phi_2$ and $\phi_3$ have the same order in $\beta$. But, as seen in Eq.~\ref{eq:Delta g3c analytic}, $\phi_2\phi_2$ dominates as the $\phi_3$ term is weighted by one more factor of $t_0^M$ decaying with OD. Hence, $g_c^{(3)}$ becomes approximately proportional to $\phi_2\phi_2$. Its spatial pattern in Fig.~\ref{fig:g3c}~(e),(f) can be explained solely from the position space behavior of $\phi_2$: $\phi_2$ is negative when two photons are nearly coincident and positive when they are at moderate separation. Hence, the product $\phi_2(x_i,x_j)\phi_2(x_j,x_k)$ is positive when three photons coincide and is negative when one is far from the other two. 

To visualize the violation from Isserlis's theorem, we calculate $\Delta g_c^{(3)}$ evaluated at zero delay in Fig.~\ref{fig:g3c}g and its temporal distribution in Fig.~S1. We observe that $\Delta g_c^{(3)}(0,0,0)$ has a sign change when OD proceeds from small to large. This arises from the competition between the positive term $t_0^{2M}$ and the negative two-photon correlated wavefunction $\sum_{i<j}\phi_2(x_i,x_j)$. At large OD, the latter dominates, reversing the sign of their sum. Since $\phi_3(0,0,0)<0$ throughout, $\Delta g_c^{(3)}(0,0,0)$ consequently changes from negative to positive. The same sign changed is simulated by the cascaded master equation with $\beta=0.06$ in Fig.~S4.

{\it Experimental considerations}---To verify the validity of our theory for experimental implementation, we consider the influence of finite power on experimental observables. For a small number of atoms ($M \leq 11$), both $g_c^{(3)}$ and $\Delta g_c^{(3)}$ can be computed numerically using the quantum regression theorem (QRT) together with the cascaded master equation. As shown in Fig.~S4, although $g_c^{(3)}$ and $\Delta g_c^{(3)}$ decrease at higher input power $P_{\rm in} \geq 0.4\Gamma_{\rm tot}$, the simulations remain in qualitative agreement with the theory for $\mathrm{OD} > 1$ (see \cite{SM}).  In experiments, the dead time in the avalanche photon detectors can influence the detection of the non-Gaussian signal from the transmitted light. For example, it reduces the measured third- and second-order correlation functions as well as the output power. Therefore, it is expected that the detector dead time will cause inaccuracy in the detection of $g_c^{(3)}$, in Fig.~\ref{fig:g3c}. We suggest that the simplest experimental signal for non-Gaussianity to measure the equal-time third order correlator $G^{(3)}(0,0,0)$ and compare its value to the prediction of Isserlis's theorem. In \cite{SM}, we model this effect using quantum-trajectory simulations and quantify the associated systematic bias and statistical uncertainty.

{\it Conclusion.}---We have developed a diagrammatic scattering theory to calculate three-photon transport through weakly coupled atomic ensembles. Our formalism provides analytic expression of the leading-order processes that contribute to non-separable three-photon correlations. We have used our new approach to predict the presence of non-Gaussian field correlations in the transmitted field of coherently driven atomic ensembles. We anticipate our results will pave the way for experimental demonstration of sources of non-Gaussian photon states using  atoms coupled to nanofibres. In future work, one could investigate whether the non-Gaussian output fields generated by atomic ensembles exhibit Wigner negativity and assess such states as potential resources for quantum computing or sensing.

{\it Acknowledgements.}---S.M. acknowledges support from the Australian Research Council (ARC) via the Future Fellowship, `Emergent many-body phenomena
in engineered quantum optical systems', project no. FT200100844. Y.W.  acknowledges the financial support from Sydney Quantum Academy, Sydney, NSW, Australia.

\bibliography{apssamp}

\end{document}